# Frequency up-conversion and arbitrary sum arithmetic of lights with orbital angular momentum

Yan Li[1,2,#1], Zhi-Yuan Zhou[1,2,#1], Dong-Sheng Ding[1,2], Wei Zhang[1,2], Shuai Shi[1,2],

Bao-Sen Shi[1,2*]

[1]*Key Laboratory of Quantum Information, University of Science and Technology of China, Hefei,*

*Anhui 230026, China*

[2]*Synergetic Innovation Center of Quantum Information & Quantum Physics, University of Science*

*and Technology of China, Hefei, Anhui 230026, China*

[*]*Corresponding author: drshi@ustc.edu.cn*

Frequency sum of two light beams carrying orbital angular momentum (OAM) in quasi-phase matching crystals was reported for the first time. The situations in which one light carried OAM and the other is in Gaussian mode and both beams carried OAM were studied in detail. An arbitrary sum arithmetic of lights with OAM was demonstrated in the conversion process. Our study is very promising in constructing hybrid OAM-based optical communication networks and all optical switching.

A light with orbital angular momentum (OAM) has stimulated great research interests since it was firstly introduced in 1992 by Allen et.al.[1], it has been used in many fields, such as optical manipulation and trapping [2-5], high precision optical measurements [6, 7], high capacity optical communications [8] and quantum information processing [9-14]. Frequency conversion is a basic technique to expand the frequency range of a fundamental light, it is widely used in optical communications [15] and up-conversion detection of infrared light [16, 17]. Frequency conversion of OAM-carried lights with nonlinear crystals (or atomic vapor) in a second (or third order) nonlinear optical process has been widely studied [18-24]. Usually, there are two kinds of phase matching methods for frequency conversion: one is birefringence phase matching (BPM), a birefringence crystal is used for realizing second harmonics generation (SHG) of a light with OAM [19]; Another is quasi-phase matching (QPM), a periodical poled crystals is used to realize SHG, which was reported very recently by our group [18]. QPM crystals have bigger effective nonlinear coefficient and no walk-off effect compared with BPM crystals. Thus QPM crystals are promising in realizing high efficient and high quality frequency conversion of an OAM-carried light [25]. For SHG of an OAM-carried light, only even values of OAM are available in SHG light, while all the OAM value is available by using sum frequency generation (SFG) as we can imprint OAM on both pump beams.

In this article, SFG of two OAM-carried lights in QPM crystals was studied for the first time. The situations when one beam carried OAM and the other beam is in Gaussian mode and both beams carried OAM were studied in detail. Arbitrary sum arithmetic of two pump beams was verified, the SFG light carried OAM that is the sum of the two pump OAM beams, which is consistent with OAM

---

conservation in the conversion process. The up-conversion of an OAM-carried in telecom band is preferable for building OAM-based up-conversion optical communication networks. For the situation when both pump beams carried OAM, the change of OAM in one pump beam affects that of the generated beam, such effect could be sued to realize all-optical mode switching, which will be applicable in mode de-multiplexing for OAM-based high capacity optical communications. The article is organized as follows: we first introduce the experimental setup; then we show the experimental results of the two above situations in detail; finally we come to the conclusion part, further perspectives of SFG of the OAM-carried lights are discussed there.

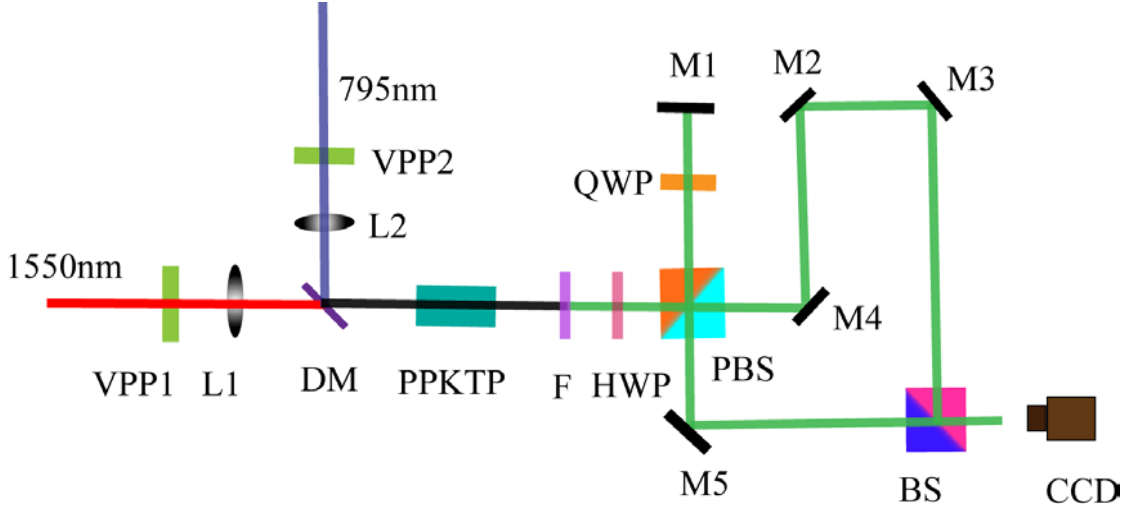

Figure 1. Experimental setup. VPP1, VPP2: vortex phase plate; L1, L2: lense; DM: dichromatic mirror; PPKTP: periodically poled KTP crystal; F: filters; HWP (QWP): half wave plate (quarter wave plate); M1-M4: mirror; PBS (BS): polarization beam splitter (beam splitter); CCD: charge coupled device camera.

The experimental setup is depicted in figure 1. The wavelengths of the two pump beams are 795 nm (from a Ti: sapphire laser, Coherent, MBR110) and 1550 nm (from a diode laser, Toptica, prodesign) respectively. Both beams are imprinted with OAM using vortex phase plates (VVPs, from RPC photonics), then are focused using lenses L1 and L2 separately before combined using a dichromatic mirror (DM). The periodically poled KTP (PPKTP, supplied by Raycol Crystals) crystal used here has dimension of 1mm×2mm×10mm, and has an poling period of 9.375 μm to get QPM for SFG of the 795 nm and 1550 nm beams to 525.5 nm, both end faces are anti-reflective coated for these wavelengths. The temperature of the crystal is controlled with a homemade semi-conductor Peilter temperature cooler with temperature stability of 2 mK. The remaining pump beams after crystal are filtered out using filters (FESH100 and FBH520-40, from Thorlabs). The polarization of the SFG light beam is rotated by a half wave plate with its optical axes placed 22.5 with respect to the vertical direction. Then the SFG light entered an interferometer to determine the OAM value of the input light. The interferometer has the same structure as our previous work [18], the function of the interferometer is to convert an input OAM light of state $|l\rangle$ into an output state of $|l\rangle + e^{i\beta}|-l\rangle$, then the output state has a flower-like spatial shape with $2l$ petals. The spatial shape of the interference pattern is monitored and acquired using a charge coupled device camera (CCD).

Before showing our experimental results, we will give some theoretical analysis firstly using the same method described in our previous study [18]. Assuming the two pump beams at 795 nm and 1550 nm are in modes $LG_0^m$ and $LG_0^n$ respectively, where *m* and *n* are the corresponding OAM value carried by the two pump beams, then the SFG light is in the mode of $LG_{(|m|+|n|-|m+n|)/2}^{m+n}$. During deviating this expression, we have assumed that the two beams match in their Gouy phase. Besides, the tight focusing and far field approximations are also fulfilled. Upon obtaining the expression of the SFG light, the experimental results can be well explained. The results when only the 1550 nm pump light carried OAM and the 795 nm pump is in Gaussian mode are showed in figure 2. The left group of images show the spatial shape of the SFG light at 525.5 nm when one arm of the interferometer is blocked, the right group of images show the interference pattern of the SFG light. The central image is the schematic layout of our VPP. For $m=0$, *n* ranges from 1 to 8, the SFG beam is in mode $LG_0^n$, therefore the OAM carried by the 1550 nm infrared light is transferred to the SFG light at 525.5 nm. The beam size of the SFG light increases with the increasing of the OAM carried by 1550 nm pump light. The numbers of petals in the right group of images are exactly in agreement with the theoretical analysis.

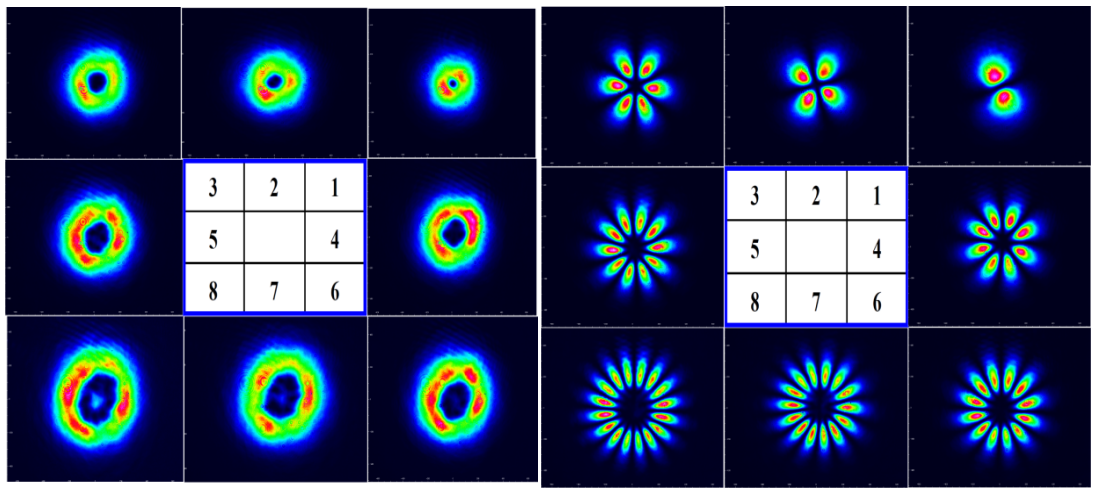

Figure 2. The experimental results when the 1550 nm pump beam carried OAM and the 795nm pump beam is in Gaussian mode. Left group of images are the spatial shapes of the SFG light when on arm of the interferometer is blocked; right group of images is the interference patterns of the interferometer of the SFG light; the central image is the schematic layout of the VPP.

The experimental results when both pump beams carried OAM are showed in figure 3. We have investigated two cases. In case one, two pump beams carried OAM with the same sign, and the 795 nm pump light carried OAM of 2, the OAM of the 1550 nm varied from 1 to 8. The results were showed in the top parts of the figure 3. The top left and the top right groups of images have the same meanings as that in figure 2. From the numbers of petals of interference patterns showed in the top right images, the SFG light carried OAM that exactly equals the sum of the two pump beams. For $m=2$, *n* ranges from 1 to 8, the SFG light has the form of $LG_0^{m+n}$. The bottom parts are the case of $m=-2$ and *n* ranging from 1 to 8, the SFG light has the form of $LG_1^{-1}$ for $n=1$ and $LG_2^{n-2}$ for $n\geq 2$. For $n=1,2,3$, we could see an out ring in the bottom left images, the intensity of the outer ring is very weak, we therefore can't see all the structures clearly. The bottom right images showed the interference patterns, the petals in the pattern showed that the OAM of the two pump beams sum to the OAM of the SFG light again. For special situation when the two pump beams carried the same value of OAM but opposite sign, there is no interference structure in the azimuthal direction (*m=-2* and *n=2*).

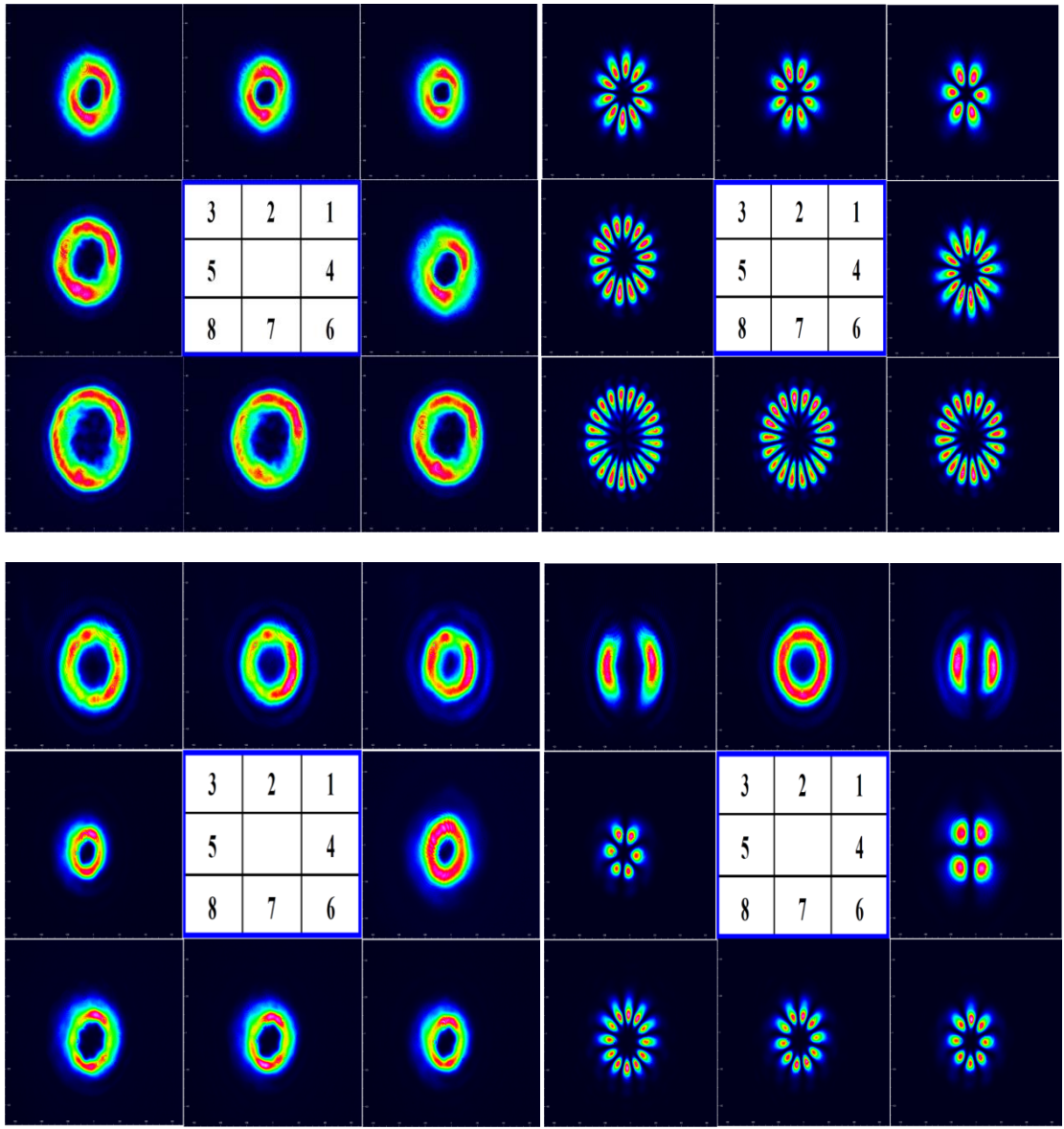

Figure 3. Experimental results when both pump beams carried OAM. Top parts for 795 nm pump beam carried OAM of 2; bottom parts for 795 nm pump beam carried OAM of -2. The top left and bottom left group of images are the spatial shapes of the SFG beams, the top right and bottom right groups of images are the corresponding interference patterns.

For summarizing, the frequency up-conversion and transfer of OAM based on SFG in QPM crystals have been studied thoroughly in this article. Arbitrary sum arithmetic of two OAM-carried pump beams is verified in the SFG process. The situations when one pump beam carried OAM and the other is in Gaussian mode and both beams carried OAM are experimentally investigated and well explained with our theoretical analysis. The present study paves the way towards high efficient and high quality engineering and processing of an OAM-carried light by applying the SFG process. The present study can be used for up-converted detecting of an infrared OAM-carried light which may be used in astrophysical observation, generating a light with OAM at specific wavelength using an OAM-carried light at wavelengths that is easy to obtain, and realizing all optical mode switching using the sum property of OAM in the SFG process. In the future, we'll put the crystal in an optical cavity and realize high efficiency up-conversion and transfer the OAM of a light.

**Acknowledgements**

This work was supported by the National Fundamental Research Program of China (Grant No. 2011CBA00200), the National Natural Science Foundation of China (Grant Nos. 11174271, 61275115, 10874171), and the Innovation Fund from the Chinese Academy of Science.